\documentclass[prb,preprint,amsmath, floatfix]{revtex4}

\usepackage{graphicx}
\usepackage{multirow}
\usepackage{dcolumn}
\usepackage{longtable}

\newcommand{\mc}{\multicolumn}

\begin{document}

\title{\textit{Ab initio} study of atomic ordering and spin-glass transition in dilute
CuMn alloys}

\author{O.~E.~Peil}
\affiliation{Condensed Matter Theory Group, Department of Physics
and Materials Science, Uppsala University, SE-75121 Uppsala, Sweden}

\author{A.~V.~Ruban}
\affiliation{Royal Institute of Technology, Department of Material
Science \& Engineering, SE-10044, Stockholm, Sweden}

\author{B.~Johansson}
\affiliation{Condensed Matter Theory Group, Department of Physics and Materials
Science, Uppsala University, SE-75121 Uppsala, Sweden} \affiliation{Royal
Institute of Technology, Department of Material Science \& Engineering,
SE-10044 Stockholm, Sweden}

\begin{abstract}
An archetypical spin-glass metallic alloy, Cu$_{0.83}$Mn$_{0.17}$, is studied
by means of an \textit{ab-initio} based approach. First-principles calculations
are employed to obtain effective chemical, strain-induced and magnetic exchange
interactions, as well as static atomic displacements, and the interactions are
subsequently used in thermodynamic simulations. It is shown that the calculated
atomic and magnetic short-range order accurately reproduces the results of
neutron-scattering experiments. In particular, it is confirmed that the alloy
exhibits a tendency toward ordering and the corresponding ordered phase is
revealed. The magnetic structure is represented by spin-spiral clusters
accompanied by weaker ferromagnetic short-range correlations. The spin-glass
transition temperature obtained in Monte Carlo simulations by a finite-size
scaling technique, 57 K, is in reasonable agreement with experimental data, 78
K.
\end{abstract}

\maketitle

\section{Introduction}

Spin-glass materials still keep on puzzling researchers with their unusual
magnetic properties. Since the beginning of 1970s, when Canella and Mydosh
\cite{cannella72} observed a cusp in the temperature dependence of the magnetic
susceptibility of a dilute solution of Fe atoms in Au, the theory of spin
glasses has become a subject of non-trivial physical and mathematical
investigations. \cite{Binder_Young_1986} Three years after the initial
discovery of the phenomenon, Edwards and Anderson \cite{edwards75} attributed
the origin of the cusp to the appearance of a spin-glass state, in which the
moments of the magnetic atoms are frozen in a disordered, glassy structure.
They showed that a substantially simplified model (EA model) could reproduce
such a state and the model became a basis for analytical mean-field
considerations. However, despite the indisputable success of the mean-field
theory to explain many of the spin-glass properties, the connection of the EA
model to real experimentally investigated spin-glass materials, such as, for
instance, dilute magnetic alloys of Mn and Fe in noble metals (Cu, Ag, and Au),
remains in most cases subtle and unclear. A great deal of failure to establish
a relation between the mean-field treatment and realistic models can be
ascribed to the lack of knowledge of the detailed structure of real materials.

Here, we present a theoretical investigation of the structure and spin-glass
behavior of Cu-Mn alloys using a realistic parameter-free model based on
first-principles calculations. There is a large amount of experimental
information on the structural and spin-glass properties of dilute
noble-metal-manganese alloys and specifically in Cu-Mn alloys.
\cite{CableW_1982, Cable_1993, Davis_Rainford_1980, Lamelas_Mydosh_1995,
Schonfeld_2004, Tsunoda_1985, Wells_Smith_1971, Werner_Gotaas_1985,
Martin_1980, Murani_1978, Gray_1982, Harders_Smith_1983, Harders_Wells_1983,
Lundgren_1983, Mezei_1982} This especially concerns the atomic and magnetic
short-range orders (ASRO and MSRO) in Cu$_{0.83}$Mn$_{0.17}$ alloys which have
been intensively investigated by various experimental techniques. In
particular, it has been established that the ASRO in Cu$_{0.83}$Mn$_{0.17}$
exhibits maxima at $Q_a=(1,1/2,0)$, \cite{CableW_1984, Roelofs_1996,
Schonfeld_2004, Tsunoda_1985, Wells_Smith_1971, Murani_1978} while the MSRO
indicates the existence of a strong tendency toward formation of magnetic
spin-density wave (SDW) clusters with a wave vector $Q_m=(1,1/2\pm\delta,0)$,
where $\delta$ is concentration dependent. \cite{Werner_Gotaas_1985,
Tsunoda_1985} In addition, static atomic displacements have just recently been
investigated by diffuse x-ray scattering in this alloy. \cite{Schonfeld_2008}
Nevertheless, the detailed atomic and magnetic structures of these alloys at
the atomic level are not known. For instance, the experimentally observed ASRO
does not allow determining unambiguously the type of atomic ordering. A
comprehensive analysis of the most accurate up-to-date experimental data
\cite{Schonfeld_2004} shows the presence of elements of several
$(1,1/2,0)$-type superstructures: \textit{D1}$_a$, \textit{DO}$_{22}$, and
Pt$_2$Mo, but none of them can be singled out as a candidate for the
low-temperature ground-state structure.

Magnetic ordering in these alloys also exhibits quite an unusual behavior.
Although the neutron-scattering experiments show strong short-range
correlations of the SDW type, no magnetic ordering occurs in the CuMn alloys at
low temperatures. Instead, typical traits of a spin glass (SG) are observed:
the cusp in the linear susceptibility, \cite{Nagata_1979} difference in the
zero-field-cooled (ZFC) and field-cooled (FC) magnetizations,
\cite{Knitter_1977} frequency dependence of the ac susceptibility,
\cite{Mulder_1981} etc. All these phenomena are connected to the SG phase
transition observed as the divergence of the non-linear susceptibility,
\cite{Levy_1988} which is equivalent to the divergence of the SG-susceptibility
(see Section~\ref{sg.section}). \cite{Binder_Young_1986}

The mechanism of the onset of the SG behavior in metallic alloys, and in the
CuMn alloy in particular, is still unclear. Along with fundamental
problems concerning the universality class of a frustrated system with
long-range interactions, there is still much to be understood in what
concerns the role of magnetic clustering in a SG transition. Indeed,
the observation of strong short-range correlations of the SDW type below and
above the transition temperature indicates that spin freezing is accompanied
by the development of magnetic clusters. There is no doubt that this
collective behavior contributes appreciably to the magnetic dynamics near the
transition. It is thus necessary to have a clear picture of the magnetic
ordering at the atomic scale.

At the same time, comprehensive theoretical investigations, in particular those
based on first-principles calculations, of atomic and magnetic orders in
Cu-rich Cu-Mn alloys are practically absent. The first {\it ab initio}
calculations of the magnetic exchange interactions in dilute Cu-Mn alloys were
performed by Oswald {\it et al.} \cite{Oswald_1985} on the basis of
Korringa-Kohn-Rostoker (KKR) impurity Green's function calculations. They
showed that the magnetic exchange interactions between Mn atoms in Cu are of
the antiferromagnetic type at the first coordination shell, but ferromagnetic
at the second and third, explaining thereby the existence of the weak
ferromagnetic SRO observed in Cu-Mn alloys.


The type of the magnetic and atomic short range order in Cu-rich Cu-Mn alloys
has also been investigated by Ling and co-workers, \cite{Ling_1994, Ling_1998}
who calculated the ASRO and paramagnetic susceptibilities in these alloys using
a so-called $S^{(2)}$-formalism with cavity corrections within density
functional theory KKR-coherent potential approximation (DFT KKR-CPA).
\cite{Gyorffy_1983} However, such a mean-filed consideration could not provide
information about the detailed atomic and magnetic structures at the atomic
level. Besides, their calculations for the Cu$_{85}$Mn$_{15}$ alloy could
reproduce neither splitting of the (1,$\frac{1}{2}$,0) magnetic peak nor the
ordering tendency for the ASRO, for which they found the phase-separation (or
clustering) behavior in direct contradiction to the experimental data.


In this work, we investigate atomic and magnetic orderings in the
Cu$_{0.83}$Mn$_{0.17}$ alloy on the basis of corresponding Monte Carlo
thermodynamics simulations, with chemical and magnetic interactions being
deduced from first-principles calculations. The outline of the paper is as
follows: Basic models, methods, and approximations, along with the description
of our approach, are presented in Section \ref{section:model}. The calculated
chemical and strain-induced interactions are analyzed in
Section~\ref{section:ecicumn}. In Section~\ref{section:asro}, the results for
the ASRO obtained from Monte Carlo simulations are compared to those provided
by experiments. Cooling down below the ordering temperature in the Monte Carlo
simulations also allows us to reveal the underlying ordered phase of a dilute
CuMn alloy. Monte Carlo magnetic simulations described in
Section~\ref{msro.section} provide us with a detailed picture of the magnetic
correlations in the system. A more thorough investigation of the SG behavior is
presented in Section \ref{sg.section}, where we show that the system exhibits a
spin-glass transition and examine the critical behavior. We conclude the
results in Section \ref{conclusion.section}.

\section{\label{section:model} Methodology}

Two different types of ordering are of interest in our case: atomic or
chemical, describing relative positions of Cu and Mn atoms on the lattice;
magnetic, associated with the spin configuration of the Mn atoms. In general,
these degrees of freedom can be interconnected in a quite complicated way due
to the strong dependence of magnetic interactions on the local chemical
environment of magnetic atoms, and owing to the dependence of the chemical
interactions on both the local and global magnetic states, as it is the case
in, e.g., FeCr alloys. \cite{Ruban_2008} However, such an interconnection is
relatively weak in Cu-rich Cu-Mn alloys. Besides, the magnetic and atomic
ordering effects are well separated in temperature.


At high temperatures relevant for the atomic local ordering, where atomic
diffusion is still possible (above 400-500 K), the alloy is in a paramagnetic
state with randomly distributed directions of local magnetic moments on Mn
atoms. The thermally induced magnetic excitations connected with the
fluctuation of the direction of spin magnetic moments on Mn atoms are several
orders of magnitude faster than the atom-vacancy interchange jumps occurring
during equilibration of the alloy. It is therefore possible to average out the
magnetic degrees of freedom and obtain effective interatomic interactions to
describe the alloy thermodynamics at temperatures substantially higher than the
spin-freezing temperature. On the other hand, at low temperatures -- of the
order of 100K and below -- where the spin-glass transition is observed, atomic
diffusion is practically absent \cite{Gray_1996} and magnetic configurations
should be determined for the Mn atoms fixed in their positions on the lattice.
Two separate problems are thus considered: finding a chemical alloy
configuration on the underlying lattice and obtaining the equilibrium ensemble
of magnetic configurations for the Mn atoms at a temperature of interest.

\subsection{\label{subsec:eci} Atomic configurational Hamiltonian
and effective cluster interactions}

The chemical, or atomic, configurational Hamiltonian used in the present work
is of an Ising type
\begin{equation}
H_{chem} = \frac{1}{2} \sum_{i \neq j} \left(V^{(2)}_{ij} + V^{\rm si}_{ij}
\right) \sigma_i
\sigma_j + \frac{1}{3} \sum_{i\neq j \neq k}V^{(3)}_{ijk} \sigma_i
\sigma_j \sigma_k + \frac{1}{4} \sum_{i\neq j\neq k \neq l}
V^{(4)}_{ijkl} \sigma_i \sigma_j \sigma_k \sigma_l ,
\label{eq:Eq1}
\end{equation}
where spin-variables, $\sigma_i$, take on values +1 or -1 if a site $i$ is
occupied by Mn or Cu atom, respectively. The effective cluster $m$-site
interactions (ECIs), $V^{(m)}_{ij\ldots k}$, and the pair strain-induced (SI)
interactions, $V^{\rm si}_{ij}$, are obtained from \textit{ab initio}
calculations as described below.

The ECI have been calculated by the screened generalized perturbation method
(SGPM) (Ref.~\onlinecite{SGPM}) on the basis of self-consistent electronic
structure calculations of random Cu-Mn alloys by the exact muffin-tin orbital
(EMTO) method \cite{Vitos} within the coherent-potential approximation (CPA).
\cite{EMTO_CPA} Since the magnetic structure stabilizes at high temperature in
a paramagnetic state, we have used a disordered local-moment (DLM)
(Ref.~\onlinecite{DLM}) magnetic configuration for Mn atoms, treating the Cu-Mn
alloy as a three-component system, Cu-Mn$^{\uparrow}$-Mn$^{\downarrow}$, where
Mn$^{\uparrow,\downarrow}$ represent spin-up and spin-down alloy components,
respectively, with the same number of Mn$^{\uparrow}$ and Mn$^{\downarrow}$
atoms distributed randomly relative to each other.

The ECIs of a binary alloy are then determined as the corresponding averages of
the ECI of the initial three-component alloy (see, for instance,
Ref.~\onlinecite{SGPM}). The intersite screening constants for the screened
Coulomb interactions contributing to the pair interactions \cite{Ruban_2002,
Ruban_2002_II} have been determined by 864-atom supercell calculations of a
random Cu$_{0.83}$Mn$_{0.17}$ alloy in the ferromagnetic state using the
locally self-consistent Green's function (LSGF) method, \cite{LSGF} assuming
that the screening constants do not depend on the magnetic configuration of Mn
atoms.


Although the size mismatch of Cu and Mn atoms is rather moderate (the atomic
volume difference of pure $\gamma$-Mn in the paramagnetic state and Cu is less
than 10\% \cite{Cowlam_1977}), and local lattice relaxations should
consequently be relatively small and give little contribution to alloy
energies, they should be included whenever a quantitative analysis is
performed, especially in case of relatively small chemical effective
interactions. In order to take the local relaxation effects into account, we
have calculated the strongest and most important strain-induced interactions at
the first three coordination shells in the dilute limit of Mn in Cu using a
supercell approach as
\begin{equation}
V^{\rm si}_{ij} = E^{(2)}_{ij,{\rm rel}} - E^{(2)}_{ij,{\rm unrel}} -
2(E^{(1)}_{{\rm rel}} - E^{(1)}_{{\rm unrel}}),
\label{eq:siint}
\end{equation}
where $E^{(1)}$ and $E^{(2)}_{ij}$ are the total energies of supercells with a
single Mn impurity and with a corresponding pair of Mn impurities in pure Cu.
Indices "rel" and "unrel" designate energies for the relaxed and unrelaxed
supercells (only atomic positions are allowed to relax, the volume and shape of
the supercells are kept fixed).

The total energies have been calculated by the projector augmented wave (PAW)
method~\cite{PAW1,PAW2} within the local density approximation (LDA)
\cite{LDA1,LDA2} as it is implemented in the Vienna \textit{ab initio}
simulation package (VASP). \cite{VASP1,VASP2,VASP3} The plane-wave cut-off
energy was set to 330 eV. If a structure has been optimized, the internal
structural parameters have been relaxed until the Hellman-Feynman forces on
each atom are less than $0.001$ eV/\AA. Two 108-, and 256-atom supercells have
been used to check the convergence with respect to the supercell size. The
presented results have been obtained for the {\it room-temperature
experimental} lattice spacing of 3.6792 \AA, \cite{Cowlam_1977} unless a
different lattice spacing is specified.

%
%
%
%
%
%

\subsection{Magnetic Hamiltonian and exchange interaction
parameters}

A Heisenberg-type Hamiltonian has been used in magnetic statistical
thermodynamics simulations
\begin{equation} \label{eq:Heis}
H = -\sum_{ij} J_{ij} c_i c_j {\bf e}_i {\bf e}_j,
\end{equation}
where ${\bf e}_i$ is a unit vector in the direction of the local magnetic
moment at a site $i$ and $J_{ij}$ the magnetic exchange interaction parameters.
The occupation numbers, $c_i$, take on values 1 or 0 if the Mn or Cu atom
occupies a site $i$, respectively, [occupation numbers are related to
$\sigma_i$ in the Hamiltonian~\eqref{eq:Eq1} as $c_i = (1 - \sigma_i)/2$]. The
occupation numbers in the Heisenberg Monte Carlo simulations have been taken
for an alloy configuration obtained from an atomic Ising Monte Carlo
simulation. The magnetic exchange interaction parameters between Mn atoms in
the alloy have been calculated using the magnetic force theorem formalism
\cite{J_xc} implemented within the EMTO method. The validity of the Heisenberg
description has been checked by comparing interactions obtained in the DLM,
ferromagnetic, and antiferromagnetic states, as well as by comparing the
results of the PAW supercell calculations with the EMTO magnetic force theorem
results, as described below.

We have also investigated the influence of the local environment effects and
ASRO on the magnetic exchange interactions for a Cu$_{0.75}$Mn$_{0.25}$ alloy.
Two supercells have been generated: one representing a completely random alloy
(the SRO parameters up to the tenth coordination shell are kept close to zero)
and another one with the SRO parameters set to equal those obtained from an
alloy Monte Carlo simulation for 300K (the SRO parameter at the first
coordination shell, $\alpha_{1} \approx -0.1$). The electronic structure of
both supercells has been calculated by the LSGF method. \cite{LSGF} The
magnetic exchange interaction parameters in these supercells have been
calculated for different pairs of Mn atoms having different numbers of Cu and
Mn atoms at the first coordination shell. The relative variation of the
nearest-neighbor exchange interaction parameter $J_{1}$ has been found to be
less than 5\% for most of the pairs. The relative difference between the
average values of $J_{1}$ for different SROs turns out to be even less
significant. The exchange interactions at larger coordination shells proved to
be almost insensitive to the local environment and SRO effects. We can thus
conclude that the fixed values of the exchange interaction parameters,
independent of the local chemical environment and global ASRO, can safely be
used in magnetic Monte Carlo simulations.

\section{\label{section:ecicumn} Effective cluster interactions in
C\lowercase{u}$_{0.83}$M\lowercase{n}$_{0.17}$ alloy}

\subsection{Chemical interactions}

\begin{figure}[htp]
\caption{(Color online) Chemical and total (chemical and strain-induced)
interactions for the first 20 coordination shells. $V^{(2)}$ and $V^{(2),{\rm bare}}$
are the chemical interactions with and without (bare) the screening contribution.
The total pair interaction in the disordered phase is given by the sum
$V^{(2)} + V^{\rm si}$.}
\label{fig3.1}
\includegraphics[width=0.8\textwidth]{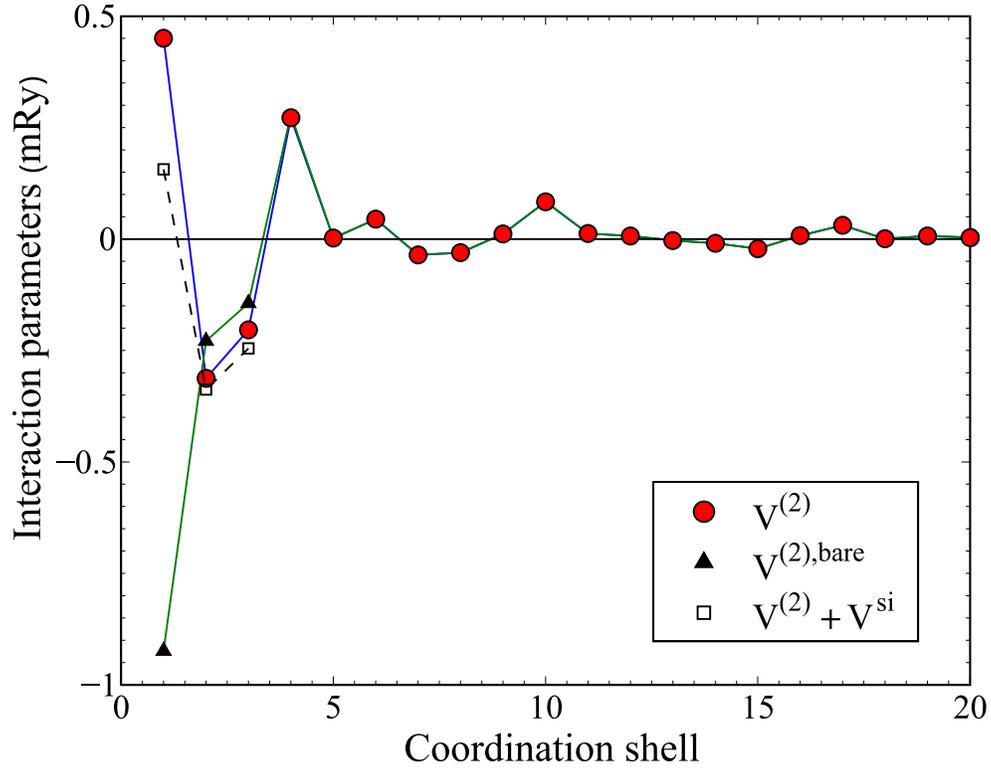}
\end{figure}

\begin{figure}[htp]
\caption{(Color online) Concentration dependence of the chemical interactions
for the first five coordination shells. The lattice constant for each concentration
is equal to the corresponding experimental value at room temperature.}
\label{fig3.2}
\includegraphics[width=0.8\textwidth]{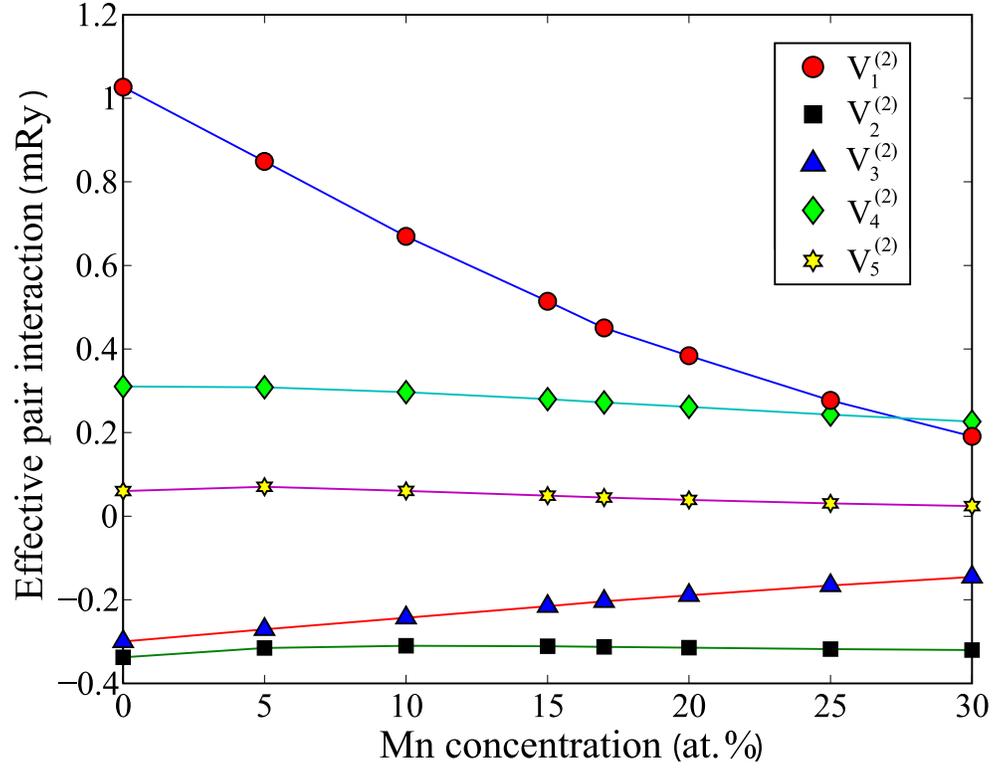}
\end{figure}

\begin{figure}[htp]
\caption{(Color online) Volume (lattice constant) dependence of the chemical
interactions for the first five coordination shells at 17 at\% Mn. The dashed
line corresponds to the room-temperature value of the lattice constant for this composition.}
\label{fig3.3}
\includegraphics[width=0.8\textwidth]{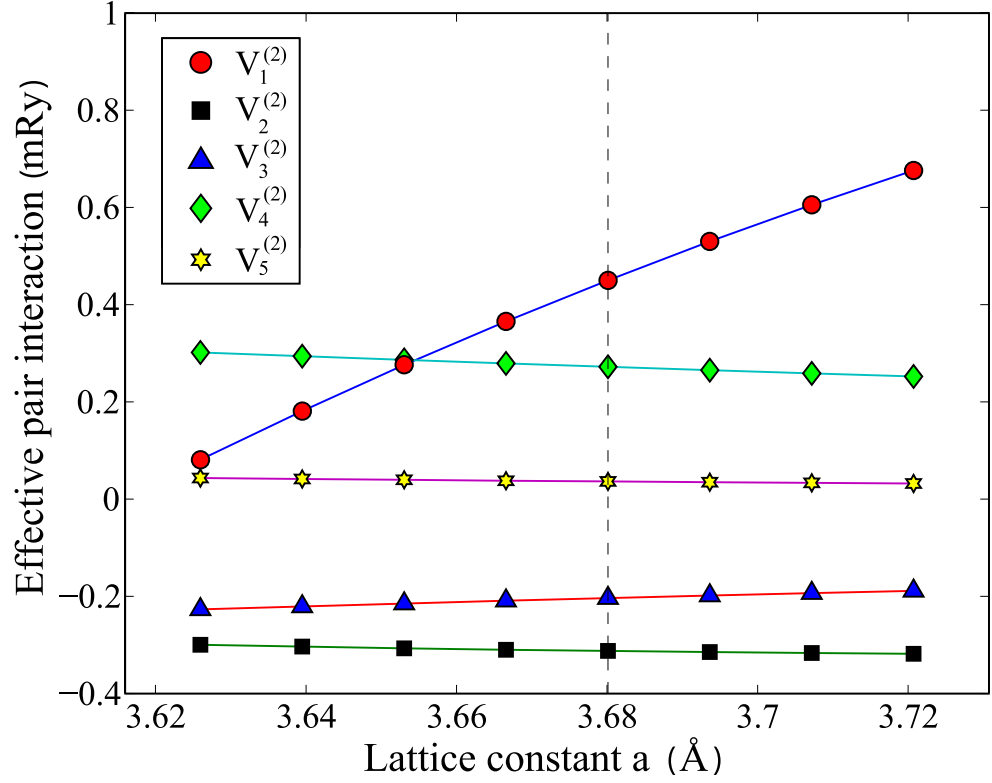}
\end{figure}

The calculated, as described in the previous section, total, chemical and,
strain-induced interactions are shown in Fig.~\ref{fig3.1}. The values of the
bare (unscreened) chemical interactions at the first two coordination shells
are shown to emphasize the importance of screening effects in the system. As
one can see, the correct account of the electrostatic contribution leads to the
change of the interaction sign at the first coordination shell. An appreciable
part of the concentration and lattice-constant (at 17 at \% Mn) dependence of
the interaction at the first coordination shell (depicted in Figs.~\ref{fig3.2}
and \ref{fig3.3}) is due to the one-electron energy contribution. The values of
the effective cluster interactions up to the 80th coordination shell, which
have been considered in the statistical thermodynamic simulations, are listed
in Table~\ref{tab:pair}.

A relatively large screening contribution to the chemical interaction at the
first coordination shell suggests that the value of the interaction is quite
sensitive to the screening parameter, which can, in fact, be quite inaccurate
due to the use of the atomic sphere approximation even with the multipole
moment correction taken into account. The accuracy of the SGPM interaction at
the first coordination shell has, therefore, been checked in supercell
calculations. This can be easily done in the dilute limit by considering, for
instance, a nearest-neighbor pair of Mn atoms in a large sample of Cu. The
total interaction energy related to the effective pair interaction in this case
is
\begin{equation}
V^{Mn-Mn}_{tot} = \frac{1}{4}\left(E^{\rm Mn-Mn} - 2E^{\rm Mn} + E^{\rm Cu}\right),
\label{eq:vtot}
\end{equation}
where $E^{\rm Mn-Mn}$ is the total energy of a supercell with two Mn atoms (in
this particular case separated by one coordination shell), $E^{\rm Mn}$ the
total energy of the same supercell but containing only a single Mn atom, and
$E^{\rm Cu}$ the total energy of pure Cu (normalized on the same supercell).
The prefactor $1/4$ is due to the definition of the Ising
Hamiltonian~\eqref{eq:Eq1}.

In a magnetic system, Eq.~\eqref{eq:vtot} leads to two different types of
interactions, depending on the relative orientation of the magnetic moments of
the Mn atoms: ferromagnetic (FM),
$V^{\textrm{Mn}\uparrow-\textrm{Mn}\uparrow}_{\rm tot}$, and antiferromagnetic
(AFM), $V^{\textrm{Mn}\uparrow-\textrm{Mn}\downarrow}_{\rm tot}$. The
interactions are calculated from the total energy of the ferromagnetic,
$E^{\textrm{Mn}\uparrow-\textrm{Mn}\uparrow}$, or antiferromagnetic,
$E^{\textrm{Mn}\uparrow-\textrm{Mn}\downarrow}$, state, respectively. However,
the interaction $V^{(2)-\mathrm{GPM}}$ is calculated in the DLM magnetic state,
and in order to compare this interaction with $V^{\rm Mn-Mn}_{\rm tot}$, proper
averaging of the latter over all possible spin configurations of the pair must
be performed. The interaction in the DLM state, relevant to our consideration
at temperatures above the spin-glass transition, is therefore given by the
following average: \cite{SGPM, Ruban_2008}
\begin{equation} \label{eq:V_sc_DLM}
V^{\rm Mn-Mn}_{\rm tot-DLM} = \frac{1}{2}\left(V^{\textrm{Mn}\uparrow-\textrm{Mn}\downarrow}_{\rm tot} +
V^{\textrm{Mn}\uparrow-\textrm{Mn}\uparrow}_{\rm tot} \right) .
\end{equation}
At the same time, the difference of these two interactions is the magnetic
exchange interaction parameter of the Heisenberg magnetic
Hamiltonian~\eqref{eq:Heis} at the corresponding coordination shell
\begin{equation} \label{eq:J_xc_sc}
J_{ij} = V^{\textrm{Mn}\uparrow-\textrm{Mn}\downarrow}_{\rm tot} -
V^{\textrm{Mn}\uparrow-\textrm{Mn}\uparrow}_{\rm tot} ,
\end{equation}
where $i$ and $j$ are the position indices of the Mn atoms in the lattice.

Besides, one should bear in mind that the interaction energy $V^{\rm
Mn-Mn}_{\rm tot}$ includes all the multi-site contributions present in the
system and they should be subtracted before a comparison with the chemical part
of the effective pair interaction is carried out
\begin{equation} \label{eq:v2sc}
V^{(2)}_{\rm sc} = V^{\rm Mn-Mn}_{\rm tot} - \frac{1}{4}\Omega^{\rm multi},
\end{equation}
where $\Omega^{\rm multi}$ is the multi-site contribution corresponding to the
right-hand side of Eq.~\eqref{eq:vtot}. It can be obtained using the GPM
multi-site interactions, calculated in the dilute limit; the multi-site
interactions are usually sufficiently accurately reproduced by the GPM since
they do not contain an electrostatic contribution. Although the multi-site
interactions are relatively weak (some of the strongest three- and four-site
interactions for the Cu$_{0.83}$Mn$_{0.17}$ alloy are listed in Tables
\ref{tab:3site} and \ref{tab:4site}) they contribute appreciably to
$\Omega^{\rm multi}$ as one can see in Table \ref{tab:scint}.

\begin{table}[h]
\caption{Interaction energies for a pair of Mn atoms in the dilute limit
obtained from the GPM ($V^{(2)-\mathrm{GPM}}(c_{\rm Mn}\rightarrow0)$) and from
supercell calculations, $V^{(2)}_{\rm sc}$. All energies are given in mRy. The
parameters $V^{\rm Mn-Mn}$ are calculated according to Eq.~\eqref{eq:vtot};
$\Omega^{\rm multi}$ is the multi-site contribution.} \label{tab:scint}
\begin{ruledtabular}
\begin{tabular}{cccccc}
$V^{(2)-\mathrm{GPM}}$ & $V^{(2)}_{\rm sc}$ & $V^{\textrm{Mn}^\uparrow-\textrm{Mn}^\downarrow}$ &
$V^{\textrm{Mn}^\uparrow-\textrm{Mn}^\uparrow}$ & $J_1$ & $\Omega^{\rm multi}$ \\
\hline
1.085 & 1.105   & --0.728 &  3.040 & --3.77  & 0.220 \\
\end{tabular}
\end{ruledtabular}
\end{table}

The total energies of 256-atom supercells,
$E^{\textrm{Mn}\uparrow-\textrm{Mn}\uparrow}$,
$E^{\textrm{Mn}\uparrow-\textrm{Mn}\downarrow}$, $E^{\textrm{Mn}}$, and
$E^{\textrm{Cu}}$ have been calculated with the PAW method (calculation details
are given in section~\ref{subsec:eci}) for the lattice spacing of 3.6792 \AA,
as has already been mentioned. These results, together with the SGPM
interaction at the first coordination shell obtained for the same lattice
spacing in the dilute limit of Mn, are presented in Table~\ref{tab:scint}. As
one can see, the agreement between the SGPM and PAW total-energy calculations
for the nearest-neighbor effective pair interaction is very good. Below, we
come back once more to the accuracy of the SGPM interactions used in the
present statistical thermodynamic simulations.

\subsection{Strain-induced interactions}

To take the effect of local lattice relaxations into account, we have
calculated strain-induced interactions at the first three coordination shells
using Eq.~\eqref{eq:siint}, as described in Section~\ref{subsec:eci}. The
interactions have been calculated in the dilute limit using the same 108- and
256-atom supercells as in the previous subsection. The results are presented in
Table~\ref{tab:siint} together with the strain-induced interactions obtained by
the Krivoglaz-Khachaturyan (KK) method \cite{Krivoglaz_1995, Khachaturyan_1983,
Blanter_1994} using the experimental value for the lattice concentration
expansion, $u = \frac{1}{a}\frac{da}{dc} = 0.0883$, for the
Cu$_{0.83}$Mn$_{0.17}$ random alloy. One can see that the KK method, as it is
implemented in Ref.~\onlinecite{Blanter_1994}, quite significantly
overestimates the interactions especially at the third and fourth coordination
shells. Therefore, in the calculations of the total effective pair interactions
we have restricted ourselves only by inclusion the strain-induced interactions
obtained in the first-principles calculations. Thus, we assume that the
contribution from more distant strain-induced interactions to the configuration
energetics is insignificant. Let us note that the long-range tail of the
strain-induced interactions given by the KK model should be overestimated
because it is valid only in the dilute limit. In random alloys, on the other
hand, the long-range tail of strain-induced interactions should actually be
damped by fluctuations of local lattice displacements due to the corresponding
fluctuations of the local chemical environment of individual atoms.

%
%
%

\begin{table}[h]
\caption{Strain-induced interactions (in mRy) obtained from the
Krivoglaz-Khachaturyan (KK) formalism and directly from supercell calculations.
"CS" stands for the coordination shell number, also given by $\alpha$. The
parameters $V^{\mathrm{si},\mathrm{KK}}_{\alpha}$ are obtained according to the
procedure described in Ref.~\onlinecite{Blanter_1994};
$V^{\mathrm{si}}_{\alpha}$ are calculated using Eq.~\eqref{eq:siint}.}
\label{tab:siint}
\begin{ruledtabular}
\begin{tabular}{cddd}
\multirow{2}*{CS, $\alpha$} &
\multirow{2}* {$V^{\mathrm{si},\mathrm{KK}}_{\alpha}$} &
\multicolumn{2}{c} {$V^{\mathrm{si}}_{\alpha}$} \\
\cline{3-4}
 & & \multicolumn{1}{c}{108-atom supercell} &
\multicolumn{1}{c}{256-atom supercell} \\
\cline{3-4}
\hline
1 &  -2.151   &  -0.992 &   -1.176 \\
2 &  -0.605   &  -0.100 &          \\
3 &   0.658   &  -0.166 &          \\
4 &   0.274   &         &          \\
5 &  -0.166   &         &          \\
6 &   0.034   &         &          \\
7 &   0.034   &         &          \\
8 &  -0.074   &         &          \\
9 &   0.091   &         &          \\
10&  -0.063   &         &          \\
\end{tabular}
\end{ruledtabular}
\end{table}

Unfortunately, there exist no reliable way of getting distant strain-induced
interactions in a random alloy from \textit{ab initio} calculations.
Nevertheless, strain-induced interactions are actually directly connected to
the corresponding static lattice displacements within the
Krivoglaz-Khachaturyan formalism. Thus, static lattice displacements in random
alloys carry important information about strain-induced interactions. To
investigate static displacements in the alloy, we have used the PAW method to
calculate the relaxed atomic positions in a 108-atom supercell containing 90 Cu
and 18 Mn atoms (corresponding to 16.7 at~\%). The chemical and magnetic
configurations have been set up in such a way as to minimize the short-range
order parameters of the ternary alloy Mn$^\uparrow$-Mn$^\downarrow$-Cu. A
static displacement for a given coordination shell has been obtained as an
average over a symmetry group corresponding to the coordination shell.

The calculated atomic displacements $\langle \bar{x}_{lmn} \rangle$ are
presented in Table~\ref{tab:sdispl}, where they are compared to the recent
experimental data by Sch\"{o}nfeld {\it et al.} \cite{Schonfeld_2008} for
Cu$_{0.83}$Mn$_{0.17}$. We also show the results for $\langle
\bar{x}^{\mathrm{CuMn}}_{lmn}\rangle$, $\langle
\bar{x}^{\mathrm{CuCu}}_{lmn}\rangle$, and $\langle
\bar{x}^{\mathrm{MnMn}}_{lmn}\rangle$ obtained from the supercell calculations
of a single Mn (256-atom supercell) and a pair of Mn atoms (108-atom
supercell), respectively. One can see that although there is a reasonable
qualitative agreement between all the calculated results and experimental data,
the quantitative differences in some cases are quite large. In particular, the
theory and experiment predict different signs for the atomic displacements of
Mn atoms in some cases. One can also note that Cu-Mn atomic displacements
calculated for the alloy supercell and for the Mn impurity are quite different.
Besides, in case of impurity they decay much slower than in case of alloy.

\begin{table}[h]
\caption{Static displacements (multiplied by $10^3$ for clarity) in units of
the lattice constant obtained from our supercell calculations and taken from
the experimental study where Georgopoulos-Cohen (GC) and $3\lambda$ separation
methods were used (see Ref.~\onlinecite{Schonfeld_2008} for details). The results
of the calculations for Mn impurity and a pair of Mn atoms at the first three
coordination shells for the values of
$\langle \bar{x}^{\mathrm{CuMn}}_{lmn}\rangle$, and
$\langle \bar{x}^{\mathrm{MnMn}}_{lmn}\rangle$, respectively, are given in
parentheses.}
\label{tab:sdispl}
\begin{ruledtabular}
\begin{tabular*}{\textwidth}{cccc|cccccc}
  & \mc{3}{c|}{Supercell calculations} &
\mc{6}{c}{Experiment \cite{Schonfeld_2008}} \\
  \cline{2-10}
 \mc{1}{c}{$lmn$} & \mc{1}{c}{$\langle \bar{x}^{\mathrm{CuCu}}_{lmn}\rangle$}
& \mc{1}{c}{$\langle \bar{x}^{\mathrm{MnMn}}_{lmn}\rangle$}
& \mc{1}{c|}{$\langle \bar{x}^{\mathrm{CuMn}}_{lmn}\rangle$}
  & \mc{2}{c}{$\langle x^{\mathrm{CuCu}}_{lmn}\rangle$}
& \mc{2}{c}{$\langle x^{\mathrm{MnMn}}_{lmn}\rangle$} & \mc{2}{c}{$\langle x^{\mathrm{CuMn}}_{lmn}\rangle$} \\
  \cline{5-10}
 &  &  &  & GC & \mc{1}{c}{$3\lambda$} & GC & \mc{1}{c}{$3\lambda$} &
GC &\mc{1}{c}{$3\lambda$} \\
\hline
110 &  -0.78 & -12.17 (-19.03) & ~3.84 ( 4.17) & -2.57 & -2.32 & -39.22 & -31.81 & ~7.20 & ~7.20 \\
200 &  ~1.03 & -3.37 (-8.06) & -2.05 (-0.39) & ~1.70 & ~1.56 & ~2.73  &  ~9.20 & -5.07 & -5.55 \\
211 &  -0.16 & ~5.33 (2.20) & ~0.83 (1.87)  & -0.07 & -0.12 & ~0.137 & ~0.21  & -0.01 & ~0.26 \\
121 &  -0.27 & -2.32 (2.83) & ~0.53 (1.27)  & -0.10 & -0.76 & ~0.9   & -0.21  & ~0.14 & ~1.94 \\
220 &  -0.57 & ~4.02  & ~0.83 (2.61)  & -0.83 & -0.70 & -11.82 & -6.07  & ~2.82 & ~2.10 \\
\end{tabular*}
\end{ruledtabular}
\end{table}

We would like to note that such a comparison should be made with caution
because the magnetic states in all these cases are different: the experiment is
done in the paramagnetic state, the impurity calculations effectively produce
results for the ferromagnetic state, and the alloy supercell calculations are
performed in a kind of a "quasi-random" magnetic state. The latter may
represent the DLM (paramagnetic) state only if the proper averaging over all
possible magnetic configurations is performed. In a single calculation, any
pair of Mn atoms is either ferromagnetically or antiferromagnetically aligned
rather than being in an average DLM state. In our view, a quite broad spectrum
of the results for static atomic displacements reflects the complexity of the
problem that requires further investigation.

\section{\label{section:asro} Atomic configuration of
C\lowercase{u}$_{0.83}$M\lowercase{n}$_{0.17}$ alloy}

\begin{figure}[htp]
\caption{Left-bottom panel: Fourier transform of the SRO parameters obtained
from Monte Carlo simulations for Cu$_{0.83}$Mn$_{0.17}$ at a temperature 500K.
Right-top panel: The atomic part of the intensity from a neutron-scattering
experiment (Ref.~\onlinecite{Schonfeld_2004}). Dark regions correspond to maxima in the
intensity; light regions, to minima. Wave vectors are given in reciprocal-lattice units (r. l. u.).}
\label{fig4.1}
\includegraphics[width=0.8\textwidth]{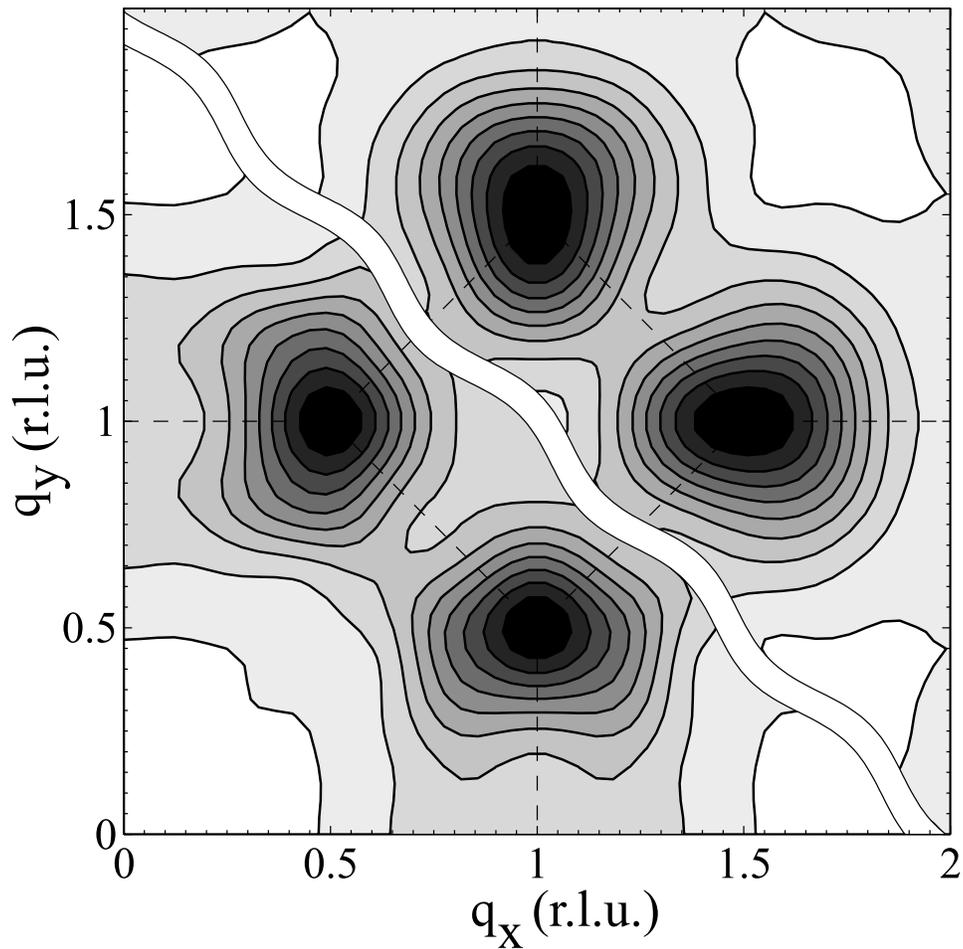}
\end{figure}

The atomic configuration of the Cu$_{0.83}$Mn$_{0.17}$ alloy has been obtained
in Monte Carlo calculations using the Metropolis algorithm in the canonical
ensemble based on the Hamiltonian~\eqref{eq:Eq1} at 500 K, which is the
annealing temperature in the diffuse scattering experiments described in
Ref.~\onlinecite{Schonfeld_2004}. Simulation boxes of different sizes (up to
$32\times 32\times 32 $ fcc elementary cubic cells) have been taken to exclude
finite-size effects. A run has consisted of 15000 sweeps, with 10000 sweeps
used for collecting statistics. The short-range order parameters have been
determined up to the 80th coordination shell; their Fourier transform is
presented in Fig.~\ref{fig4.1} along with the experimental data from
Ref.~\onlinecite{Schonfeld_2004}. Clearly, the ASRO is reproduced very well up
to the width of the peaks. A slight difference in the shape can be either due
to a small error originating from the cut-off of the interaction range, and
multi-site terms for more than 4 sites, or due to an incomplete subtraction of
the magnetic contribution from neutron-scattering intensity data. The latter
claim is supported by subtracting the calculated normalized atomic intensity
pattern from the experimental one and noting that their difference matches the
magnetic intensity pattern quite well.

\begin{figure}[htp]
\caption{(Color online) Ordered structure $DO_{60}$. Light spheres represent
the Mn atoms; dark, the Cu atoms. Smaller and larger spheres depict,
respectively, positions at the corners and at the face centers of an elementary
cubic cell. The elementary cell of the ordered structure is marked with a frame.}
\label{fig4.2}
\includegraphics[width=0.8\textwidth]{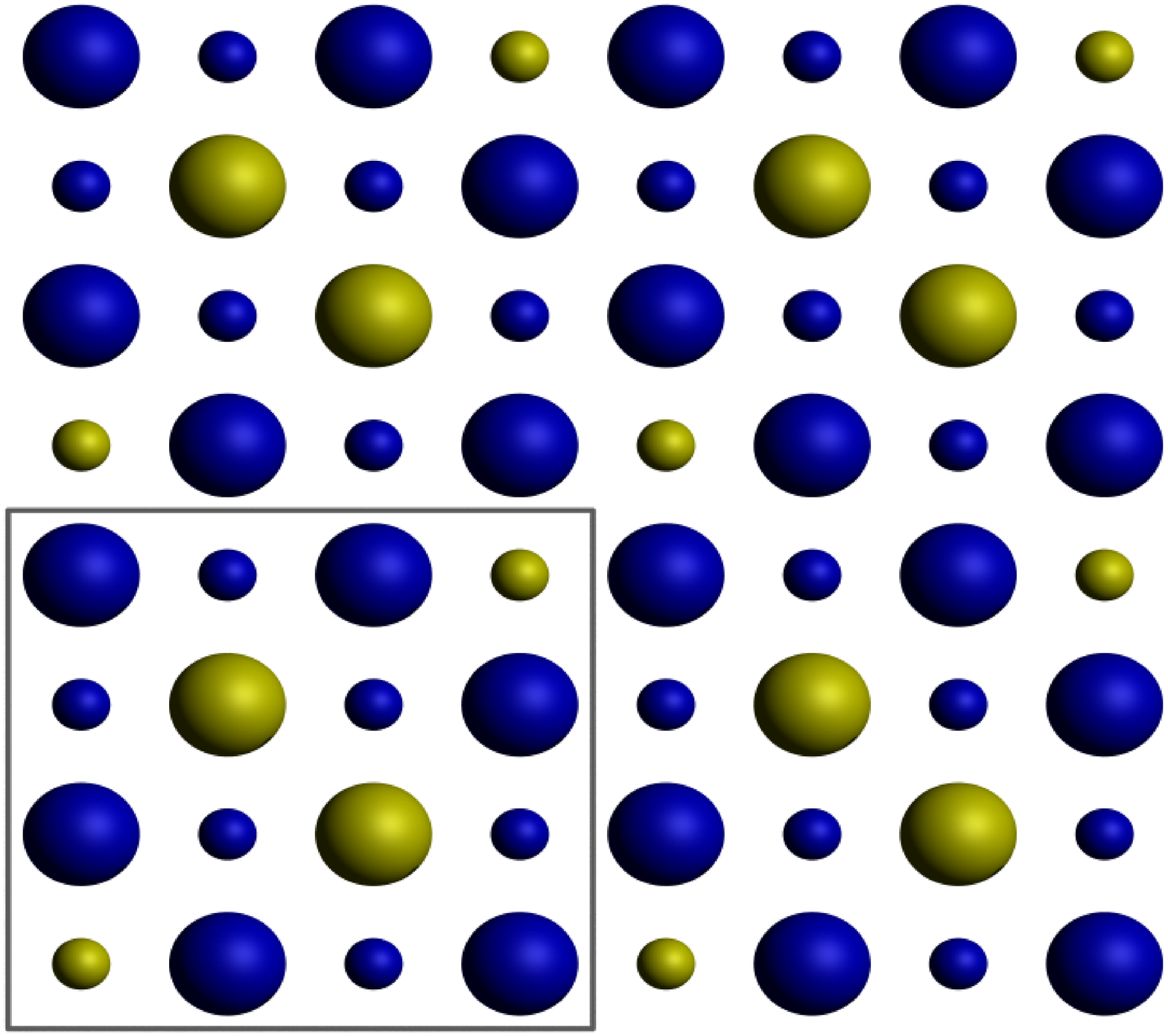}
\end{figure}

The ASRO in the Cu$_{0.83}$Mn$_{0.17}$ alloy is characterized by
$Q_a=(1,1/2,0)$ type of atomic ordering.  The local maximum of the Fourier
transform of the SRO parameters at $(0,0,0)$ is also present, yet it is very
weak, implying a strong tendency toward ordering rather than segregation. There
has been a controversy as to what kind of an ordered phase this ASRO
represents. \cite{Tsunoda_1992, CableW_1984, Hirabayashi_1978, Bouchiat_1981}
None of the previously proposed ordered structures (e.g., $DO_{22}$, $A_2B_2$,
$D1_a$, etc.) satisfies all the necessary conditions, such as the absence of
additional reflections at points inequivalent to $Q_a$ (such as for $DO_{22}$)
or the abundance of Cu in the ordered precipitates (this disfavors $A_2B_2$).
To find the sought-for ordered structure, we have performed Monte Carlo
simulations down to low temperatures. Below the transition temperature ($T_c
\approx 250K$), an ordered phase with 25 at\% Mn has precipitated. The
structure of the phase (first proposed in Ref.~\onlinecite{kulkarni88} as an
ordered phase for the Ni$_3$Mo alloy) is shown in Fig.~\ref{fig4.2}. It
consists of stripes similar to the $A_2B_2$ structure, but with half of the Mn
stripes being replaced by Cu.

It is worth noting, however, that the ordering energy of the obtained Cu$_3$Mn
phase (referred to as $DO_{60}$ hereafter) is very close to that of the $D1_a$
(Ni$_4$Mo) phase. It turns out that three-site chemical interactions play a
crucial role in stabilizing the $DO_{60}$ structure. Unfortunately, there is no
simple way to check the stability of the Cu$_3$Mn and Cu$_4$Mn phases with
respect to each other, since this would require calculating the free energy for
the alloy at different concentrations, which is a formidable task due to a
strong effect of magnetism on the alloy thermodynamics.

\begin{table}
  \caption{Ordering energies of different Cu$_3$Mn structures: $E_{\rm tot}$
are calculated as the difference between the total energy of an ordered phase
and a random alloy, with the Mn atoms being in the DLM state;
$E_{\rm SGPM}$ are obtained from the effective interactions $V^{(m)}$
($m = 2,3,4$).}
  \begin{ruledtabular}
  \begin{tabular}{ldddd}
  Structure & \mc{1}{c}{E$_{\rm SGPM}$ (mRy)} & \mc{1}{c}{E$_{\rm tot}$ (mRy)} &
              \mc{1}{c}{E$_{\rm SGPM}$ - E$_{\rm SGPM}^{ DO_{60}}$ (mRy)}&
              \mc{1}{c}{E$_{\rm tot}$ - E$_{\rm tot}^{ DO_{60}}$ (mRy)} \\
  \hline
  L$_{12}$  &  1.68 &  2.6  & 2.51 & 2.89 \\
  DO$_{22}$ & -0.83 &  0.26 & 0.54 & 0.54 \\
  DO$_{60}$ & -1.68 & -0.29 & 0.0  & 0.0  \\
  \end{tabular}
  \end{ruledtabular}
  \label{tab.eord}
\end{table}

The ordering energies of the obtained $DO_{60}$ phase, as well as the energies
of some other ordered structures for the same composition, have been calculated
and compared to direct total-energy calculations performed with the EMTO
method. The results are presented in Table~\ref{tab.eord}. It is obvious that
among the other Cu$_3$Mn phases, the $DO_{60}$ phase is the most stable one at
the given composition, and the energy difference with respect to the DO$_{22}$
structure ensures that minor corrections (static displacements, phonon
contribution, etc.) cannot change the observed behavior.

\section{Magnetic structure}
\label{msro.section}

As a rule, accurate numerical studies of SG systems by Monte Carlo methods are
computationally very demanding. However, a general picture of magnetic
correlations turns out to be relatively insensitive to the quality of sampling.
We have employed the Metropolis algorithm for the Hiesenberg Monte Carlo
simulations to obtain pair spin-spin correlations in the temperature range
between 300 and 20K. The atomic configurations have been obtained in the Ising
atomic Monte Carlo simulations at 500 K and kept fixed during magnetic
simulations because the atomic diffusion is practically absent at ambient and
lower temperatures.

Despite frustration that impedes equilibration in simulations even above the SG
transition temperature, the spin-spin correlations tend to saturate to their
equilibrium values much faster than quantities characterizing the SG phase. To
put this another way, the autocorrelation time associated with the spin-spin
correlations changes smoothly and remains relatively small down to the lowest
temperatures owing to the absence of the second-order transition. We can thus
make do with a simplified Monte Carlo technique to achieve a reasonable
accuracy with a moderate number of time steps in this case.

\begin{figure}[htp]
\caption{Left-bottom panel: Fourier transform of the spin-spin correlation
function obtained from Monte Carlo simulations for Cu$_{0.83}$Mn$_{0.17}$ at
a temperature 20K. Right-top panel: The magnetic part of the intensity from
a neutron-scattering experiment (Ref~\onlinecite{Schonfeld_2004}). Dark regions correspond
to maxima in the intensity; light regions, to minima. Wave vectors are given in
reciprocal lattice units (r. l. u.).}
\label{fig5.1}
\includegraphics[width=0.8\textwidth]{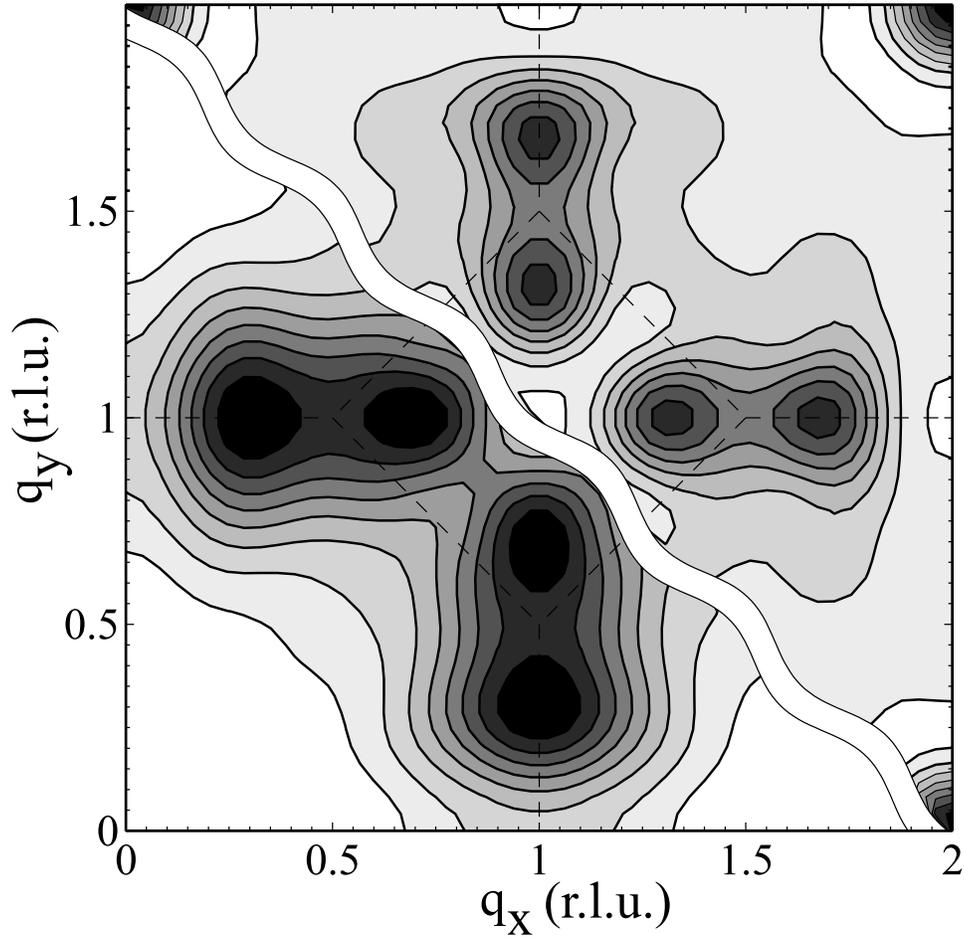}
\end{figure}

To characterize a MSRO, a spin-spin correlation function
$C(R)=\frac{1}{N}\sum_i\langle \vec{S}_i\vec{S}_{i+R}\rangle$ has been
calculated. The Fourier transform of the spin-spin correlation function
obtained from the simulations at 20K is presented in Fig.~\ref{fig5.1} along
with the experimental data on the corresponding polarized neutron-scattering
intensities. The maxima at $Q_s=(1,1/2\pm\delta,0)$ ($\delta\approx0.21$ for
$c_{\rm Mn}=0.172$) correspond to incommensurate SDW clusters. It has already
been contemplated \cite{Tsunoda_1992} that the local magnetic order represents
a non-collinear single-$Q$ spin-spiral structure, with the directions of
propagation of SDWs varying in different domains. This type of the local
magnetic order is observed in simulations at low temperatures (20K). Since the
correlations at the first several coordination shells are strong, these
clusters, rather than individual spins, determine the spin dynamics close to
the freezing temperature $T_{\rm SG}$. Also, non-collinear MSRO clusters may be
responsible for the chirality ordering that accompanies the SG transition.
\cite{Kawamura_1992}

It is worth noting that we neglect anisotropy for the pure
Cu$_{0.83}$Mn$_{0.17}$ alloy because the anisotropy energy for this alloy is
too small (of the order of $10^{-3}$ meV per Mn atom \cite{Kouvel_1961}) to
have any considerable effect on the MSRO. A situation, however, can be
different in CuMn alloys containing heavy impurities. For instance, adding even
a small amount of Pt to the alloy results in the anisotropy energy of up to
$0.1$ meV per Mn atom. \cite{Fert_1980, Prejean_1980}

\section{Spin-glass behavior}
\label{sg.section}

In simulating a SG system, one has to take into account the fact that most of
the quantities specific to a SG state, such as SG susceptibility, overlap
parameters, etc., are quite sensitive to a realization of disorder
\cite{Binder_Young_1986} and a configurational averaging must therefore be
carried out. The latter requires a large number of independent runs. Moreover,
to obtain unbiased values of ensemble averages for each configuration, it is
important to have samples in the equilibrium state. However, due to magnetic
frustrations in the alloy it is hard to achieve the equilibration, and improved
methods must be employed to get results on a reasonable time scale.

Slow equilibration imposes a restriction on a system size in simulations and
therefore finite-size effects become an important issue. To overcome the size
and boundary dependences, we resort to a technique that has become a standard
tool in investigations of critical phenomena, namely, the finite-size scaling
procedure. \cite{Barber_1984, Landau_2000} The general idea is to obtain a
dimensionless correlation length, $\xi/L$, for different system linear sizes,
$L$ (defined in terms of the number of elementary cubic cells of the fcc
lattice), and use the scaling law
\begin{equation}
\xi_L/L = f(L^{1/\nu}(T-T_c)),
\label{eq:cl_fss}
\end{equation}
to determine a transition temperature $T_c$ by taking advantage of a simple
corollary of the fact that the dimensionless correlation length becomes
independent of $L$ at $T_c$. To calculate the correlation length, two
independent replicas with identical atomic distributions have been simulated in
parallel and the correlation length has been evaluated according to the
following well-known formula: \cite{Lee_Young_2003}
\begin{equation}
\xi_{L} = \frac{1}{2\sin{(k^x_{\rm
min}/2)}}\sqrt{\frac{\chi_{\rm SG}(\textbf{0})} {\chi_{\rm SG}({\bf
k}_{\rm min})}-1},
\end{equation}
where $\textbf{k}_{\rm min} = 2\pi/L(1,0,0)$ and the SG correlation function
$\chi_{\rm SG}(\textbf{k})$ is defined as
\begin{equation}
\chi_{\rm SG}(\textbf{k}) = N \left[\sum_{\mu\nu}\left\langle
\left|q^{\mu\nu}(\textbf{k}) \right|^2\right\rangle\right]_{\rm av},
\end{equation}
\begin{equation}
q^{\mu\nu}(\textbf{k}) = \frac{1}{N}\sum_i
S_i^{\mu(1)}S_i^{\nu(2)}e^{i{\bf k}{\bf R}_i},
\end{equation}
with indices $(1)$ and $(2)$ designating quantities related to the two
replicas; summations are performed over all magnetic atoms in the alloy,
$\langle\dots\rangle$ and $[\dots]_{\rm av}$ stand for thermal and
configurational averaging, respectively, and $N=4L^3$. The FSS scaling of the
SG susceptibility $\chi_{\rm SG} \equiv \chi_{\rm SG}(\textbf{0})$ is given by
the relation
\begin{equation}
\chi_{\rm SG}/L^{2-\eta} = X(L^{1/\nu}(T-T_c)).
\end{equation}

Monte Carlo simulations have been performed using the heat-bath algorithm.
\cite{Olive_1986} To make calculations practically affordable, only the
exchange interaction parameters with $|\textbf{R}_{ij}| < r_0 = 4$ elementary
cubic cells have been taken into account. To ensure thermalization, logarithmic
binning has been applied to the spin-glass susceptibility $\chi_{\rm SG}(0)$
and a consistency check of the obtained configurational averages has been made
in a way similar to that in Ref.~\onlinecite{Bhatt_Young_1985}, namely, the
following two quantities have been calculated:
\begin{equation}
\chi_{o} = \frac{1}{N}\left[\sum_i\left\langle
\textbf{S}_i^{(1)} \cdot \textbf{S}_i^{(2)}\right\rangle^2\right]_{\rm av}
\end{equation}
and
\begin{equation}
\chi_{d} = \frac{1}{T}\sum_t\left[\frac{1}{N}\left(\sum_i
\textbf{S}_i^{(1)}(t_0) \cdot \textbf{S}_i^{(1)}(t_0+t)\right)^2\right]_{\rm
av},
\end{equation}
where $t$ designates the time in terms of Monte Carlo steps, $t_0$
is the equilibration time, and $T$  the size of the time
window during which measurements are carried out. A calculation is
considered converged with respect to the configurational averaging if a
condition $\chi_{o}=\chi_{d}$ is satisfied with a high accuracy
(to within 1\%). To accelerate equilibration, we have used the
overrelaxation method,  \cite{Campos_2006} which consists in
performing microcanonical steps; that is, all spins are flipped
sequentially according to the following energy-conserving transformation:
\begin{equation}
\textbf{S}_i \rightarrow \textbf{S}_i -
\frac{2\left(\textbf{S}_i\cdot\frac{\partial H}
{\partial\textbf{S}_i}\right)\frac{\partial H}{\partial
\textbf{S}_i}}{\left(\frac{\partial H}{\partial
\textbf{S}_i}\right)^2}.
\end{equation}
$L$ sequential microcanonical steps for each heat-bath sweep have been found to
be optimal in terms of the actual calculation time.

\begin{figure}[htp]
\caption{(Color online) Finite-size scaling of the SG susceptibility.}
\label{fig6.1}
\includegraphics[width=0.8\textwidth]{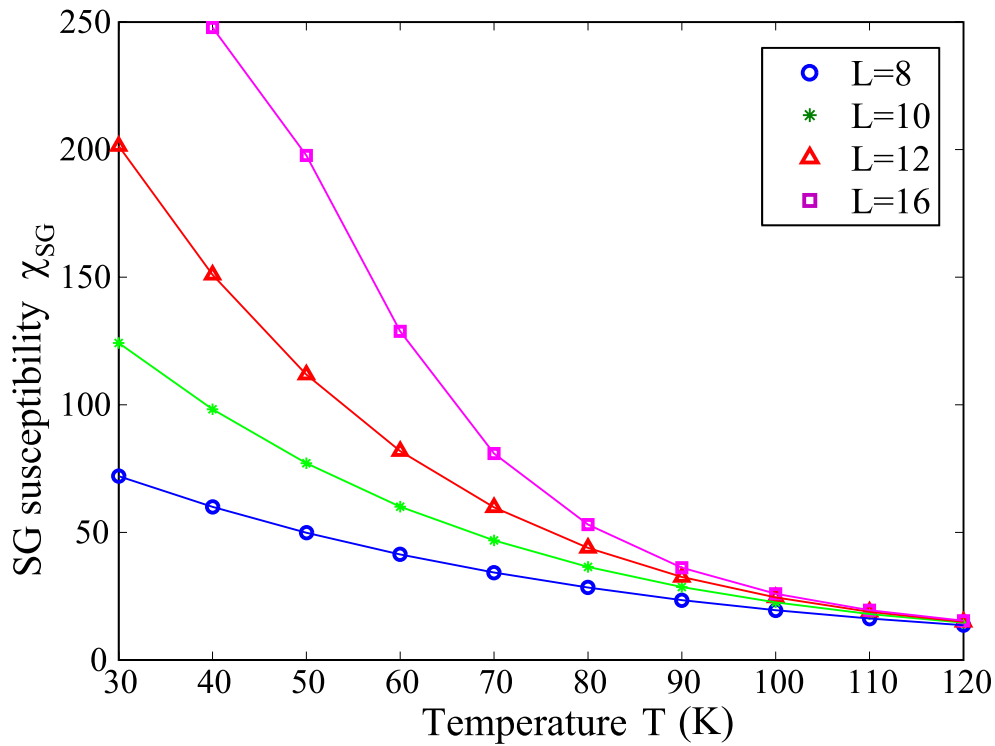}
\end{figure}

\begin{figure}[htp]
\caption{(Color online) Finite-size scaling of the SG correlation length.
Curves for different system sizes cross at $T_c=57\pm 5$K.}
\label{fig6.2}
\includegraphics[width=0.8\textwidth]{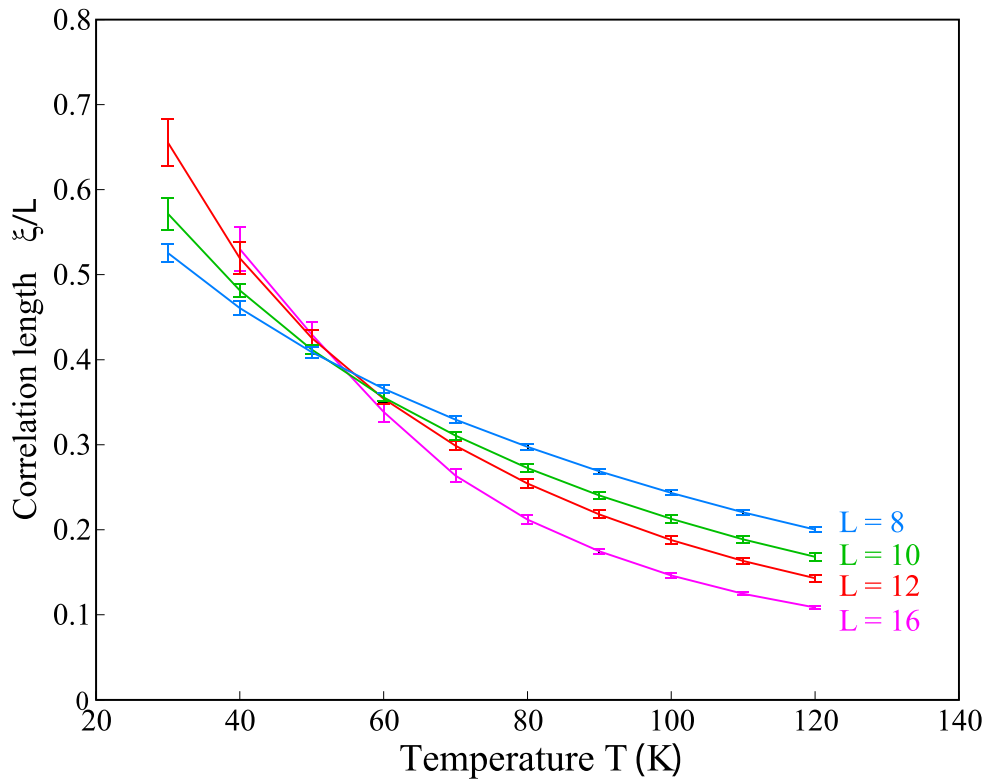}
\end{figure}

\begin{figure}[htp]
\caption{(Color online) Universal behavior of the SG correlation length.
Best fit is achieved with $\nu=0.95$.}
\label{fig6.3}
\includegraphics[width=0.8\textwidth]{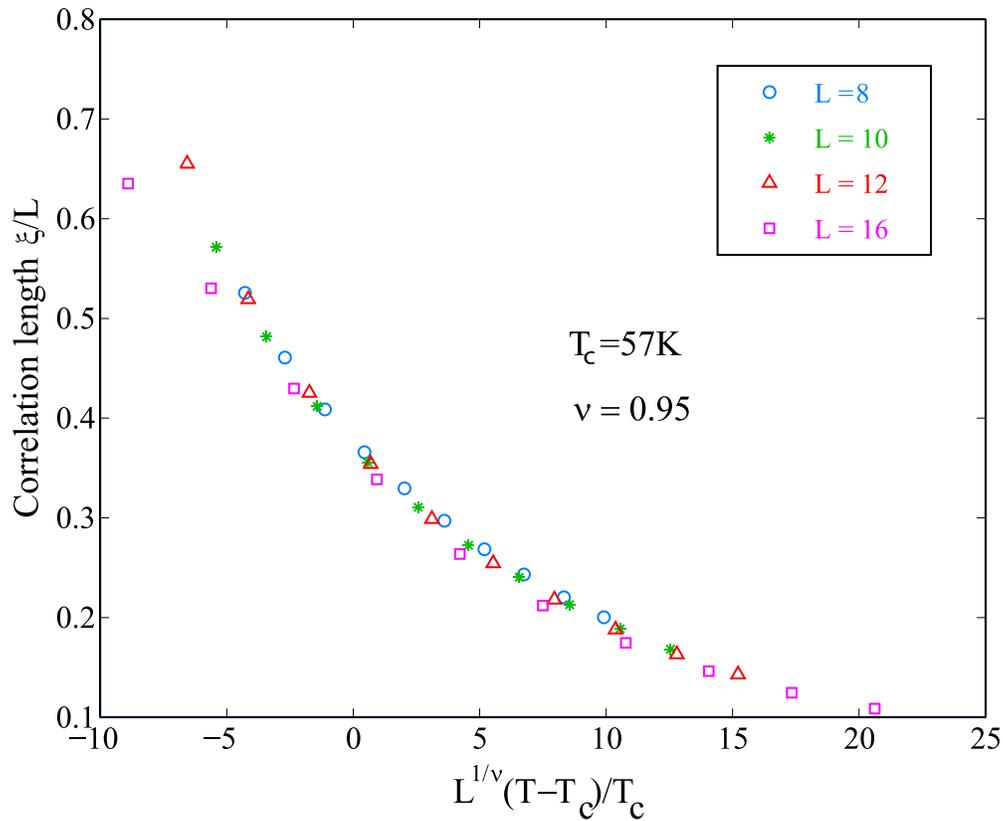}
\end{figure}

\begin{figure}[htp]
\caption{(Color online) Universal behavior of the SG susceptibility.
Best fit is achieved with $\eta=0.25$.}
\label{fig6.4}
\includegraphics[width=0.8\textwidth]{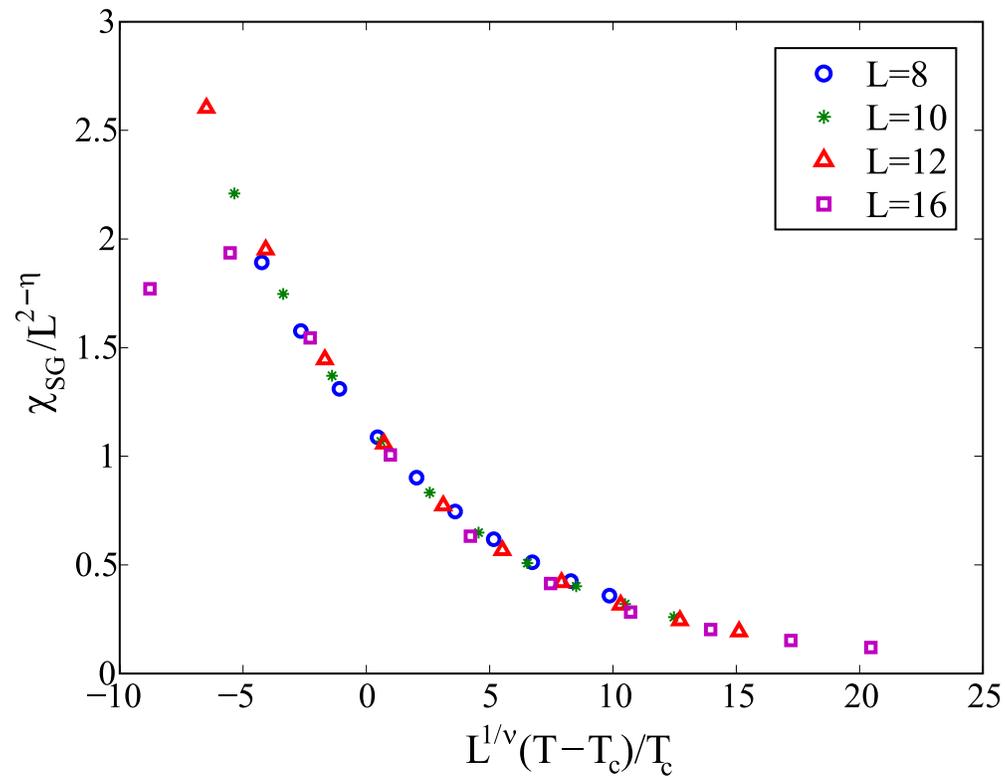}
\end{figure}

The calculated size dependence of the SG susceptibility (Fig.~\ref{fig6.1})
suggests the divergent behavior below 100K. More detailed information can be
extracted from the size scaling of the dependence of the correlation length on
temperature (Fig.~\ref{fig6.2}). The results show unequivocally that the system
undergoes a transition at a finite temperature. The absence of a long-range
order has been checked by observing the development of the correlation length
corresponding to the SDW ordering. Although the SG transition is clearly
observed, one cannot rule out a crossover to a marginal behavior at larger
supercell sizes. The data exhibit a universal scaling with $T_c = 57 \pm 5K$
and $\nu=0.95 \pm 0.1$ (Fig.~\ref{fig6.3}). Using these parameters, the SG
susceptibility can also be fitted providing a parameter $\eta=0.25 \pm 0.1$
(Fig.~\ref{fig6.4}). From the relation $\gamma=\nu (2-\eta)$, one can also find
that $\gamma\approx 1.7 \pm 0.3$. In all plots, the deviation of the data for
$L=16$ is due to the systematic underestimation of the SG susceptibility at low
temperatures.

\begin{figure}[htp]
\caption{(Color online) The extended scaling of the SG correlation length according to Eq.~\eqref{eq:extscal}.
The reduced temperature $t$ is defined as $t=T/T_c$.}
\label{fig6.5}
\includegraphics[width=0.8\textwidth]{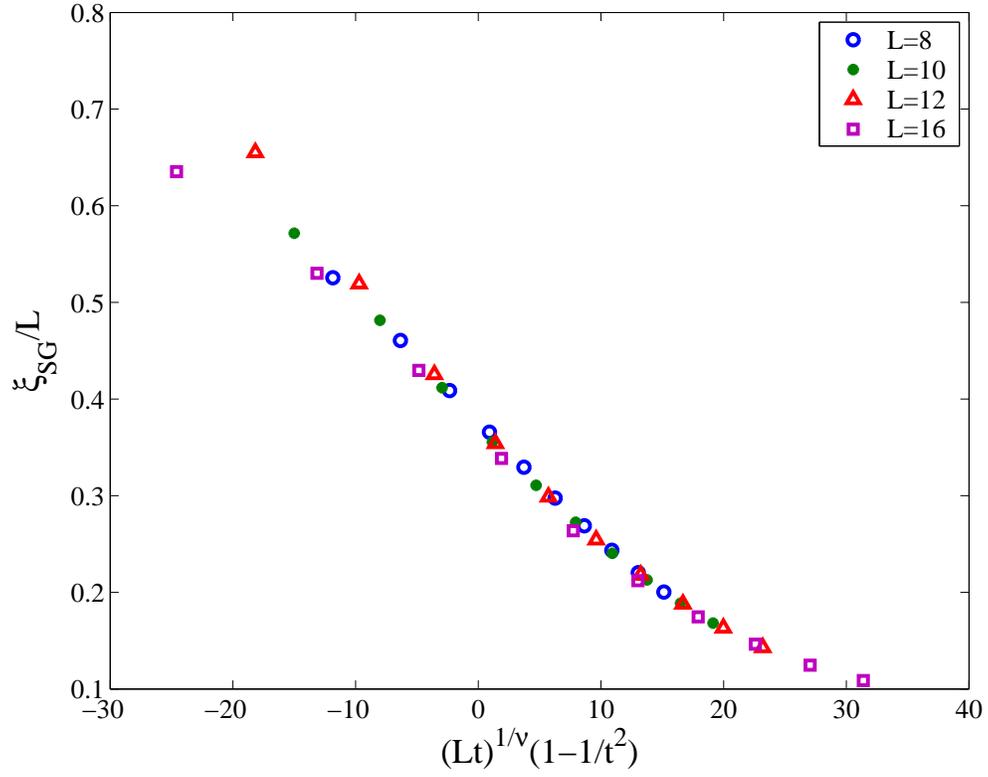}
\end{figure}

\begin{figure}[htp]
\caption{(Color online) The extended scaling of the SG susceptibility according to Eq.~\eqref{eq:extscal}.
The reduced temperature $t$ is defined as $t=T/T_c$.}
\label{fig6.6}
\includegraphics[width=0.8\textwidth]{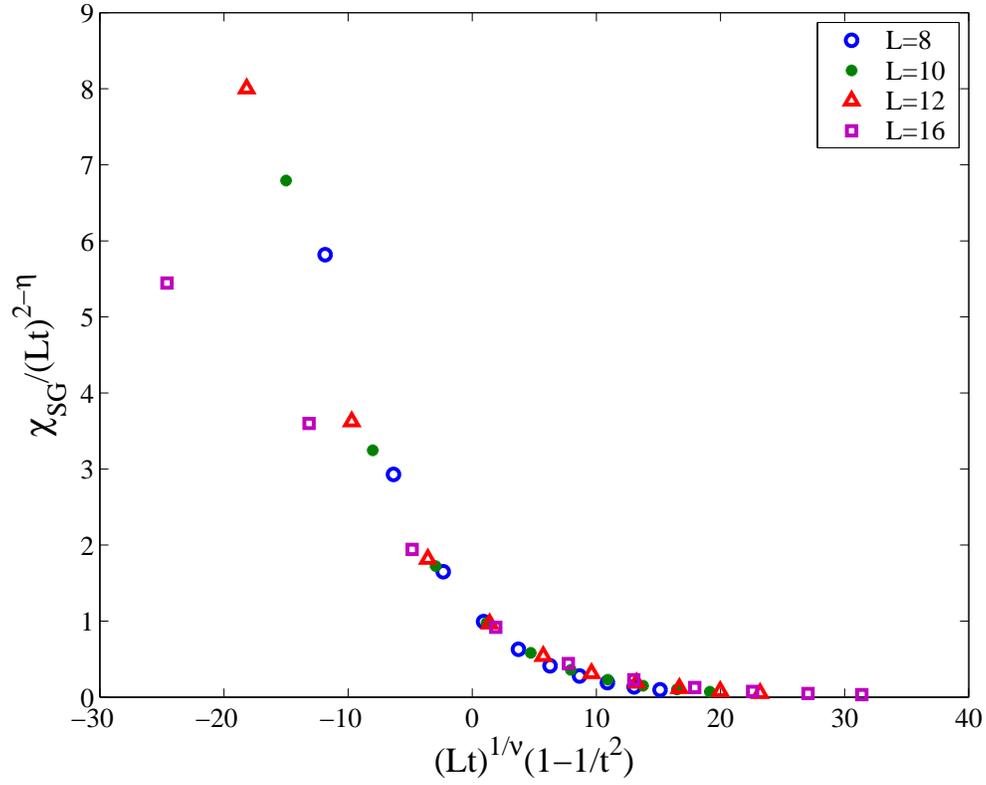}
\end{figure}

An alternative (extended) scaling relation based on the high-temperature series
expansion was proposed by Campbell \textit{et al}. \cite{Campbell_2006} Unlike
the conventional scaling [Eq.~\eqref{eq:cl_fss}], the extended scaling applies
not only to a close vicinity of the critical region but must also hold for all
temperatures above the transition. The extended relation, that can be written
as \cite{Campbell_2006}
\begin{align}
\xi_L/L = & \tilde{f}\left((LT)^{1/\nu}\left[1-\left(T_c/T\right)^2\right]\right), \\
\chi_{SG}/(LT)^{2-\eta} = & \tilde{X}\left((LT)^{1/\nu}\left[1-\left(T_c/T\right)^2\right]\right),
\label{eq:extscal}
\end{align}
has been applied to the calculated data. The results shown in
Figs.~\ref{fig6.5} and \ref{fig6.6} demonstrate a rather good scaling, although
a larger temperature interval and higher accuracy are required to make a more
thorough comparison of the two types of scalings.

The values of the critical exponents differ from those obtained in experimental
measurements \cite{Levy_1988} from which we have $\nu_{exp}=1.3 \pm 0.15$ and
$\gamma_{exp}=2.3 \pm 0.2$. One of the possible causes of this disagreement is
a rather high sensitivity of the critical exponents to the scaling corrections
which might be non-negligible in our calculations because the range of the
interactions is comparable to the size of the simulation box. Another effect,
which can strongly influence the critical behavior, is anisotropy. It is well
known that even a small amount of anisotropy present in a real system can
change the critical behavior close to the transition temperature, giving rise
to the values of critical exponents different from those for an isotropic
system. Unfortunately, it is practically impossible to detect any crossover
from an isotropic to anisotropic critical behavior in small systems such as the
one we have used here and with such a small anisotropy energy typical for pure
CuMn alloys. Simulations for systems with an appreciable amount of anisotropy
(such as, e.g., CuMn$_x$Pt$_y$) can therefore be an interesting problem for
future work.

The obtained transition temperature, 57K, turns out be slightly lower than the
experimental value (~78K). One of the possible explanations for this
discrepancy is the cut-off of the interaction range that we had to introduce.
On the other hand, Monte Carlo dynamics (only local updates without
overrelaxation) has revealed a strong critical slowing down at about a
temperature equal to 80K, and this value only slightly depends on the cut-off.
This could indicate that while dynamics is driven primarily by short-range
interactions, the SG transition itself is very sensitive to long-range
interactions which may play a rather important role in the onset of the SG
phase in the CuMn alloys.

In spite of a possible strong influence of long-range interactions on the SG
dynamics, it seems unlikely that the dominating contribution to the onset of
the transition comes from effectively infinite-range interactions because a
natural cut-off can be considered originating from various kinds of
imperfections (vacancies, impurities, static displacements, etc.) present in
any real material but discarded in the current study. For instance, we have
neglected static atomic displacements in the magnetic part of the problem.
Although their effect should be very small on the effective interactions at
nearest-neighbor interactions, more distant interaction will be definitely
damped because of the corresponding exponential damping of the Cu $s$-like
states. This means that such a system will doubtfully cross over to a
mean-field-like transition in a realistic model.

\section{Concluding remarks}
\label{conclusion.section}

The basic result of this work is the structure of the CuMn alloy obtained from
first-principles calculations. It has been shown that both the atomic and
magnetic thermodynamics simulations accurately reproduce the corresponding
experimental data. Interestingly, it has already been pointed out in some
studies \cite{CableW_1984} that the most probable stoichiometry of the ordered
phase of a Cu-rich CuMn alloy is Cu-25 at\% Mn; however, the scattering pattern
of known structures with this composition ($DO_{22}$, $L_{12}$) is not
compatible with the observed one, and the $A_{2}B_{2}$ structure, that lacks
this drawback, has been considered as a best description of the ASRO. In this
respect, the structure $DO_{60}$, on one hand, contains 25 at\% Mn and, on the
other hand, resembles the $A_2B_2$ structure closely and produces a similar
scattering intensity picture with the dominant peaks at $Q_a=(1,1/2,0)$. This
reconciles a seeming discrepancy between the observed stoichiometry and the
structure of the ASRO.

In spite of some limitations of the approach used, a large class of metallic
spin-glass materials, e.g., AgMn, AuMn, PtMn, AuFe, etc., can be investigated
in a similar fashion. Other contributions to the magnetic interactions might be
required to be taken into consideration, though. Although in the CuMn alloy
relativistic effects can be neglected, in systems with heavy elements (Au, Pt)
they are important and can lead to strongly anisotropic interactions. Most of
the anisotropic contributions can be treated within the GPM formalism in the
framework of a fully relativistic code (see, e.g.,
Ref.~\onlinecite{Pourovskii_2005}), extending thus the approach to highly
anisotropic heavy-element metallic alloys.

\acknowledgments { We are grateful to the Swedish Research Council (VR) and the
Swedish Foundation for Strategic Research (SSF) for financial support.
Calculations were done using UPPMAX (Uppsala) and NSC (Link\"oping) resources.}

\appendix*
\section*{\label{section:app} Appendix}

Effective cluster interactions for two-site, three-site and four-site clusters
are presented in Tables~\ref{tab:pair}-\ref{tab:4site}. In all cases, an
$m$-site cluster is represented by the first atom at the origin, $(0,0,0)$, and
$(m-1)$ vectors corresponding to the other atoms in the cluster as given in the
tables. The interactions are given in millirydberg (mRy). Note that the
accuracy of the EMTO calculations has been set to $10^{-3}$ mRy.


\pagebreak
\begingroup
\squeezetable

\begin{table*}
\centering \caption{Pair effective and magnetic exchange
interactions for the first 80 coordination spheres.}
\label{tab:pair}
\begin{ruledtabular}
\begin{tabular*}{0.9\textwidth}{@{\extracolsep{\fill}}ccddc|ccdd}

N & \mc{1}{c}{$(R_x,R_y,R_z)$} &\mc{1}{c}{$V^{(2)}_{(R_0,R)}$}
& \mc{1}{c}{$J^{(xc)}_{(R_0,R)}$} & &
N & \mc{1}{c}{$(R_x,R_y,R_z)$}  & \mc{1}{c}{$V^{(2)}_{(R_0,R)}$}
& \mc{1}{c}{$J^{(xc)}_{(R_0,R)}$}\\
 \hline
%
%
%

 1 & (0.5,~0.5,~0.0) & ~0.45035 & -2.51559 & & 41 & (2.5,~2.5,~2.0) & ~0.00003 & -0.00029 \\
 2 & (1.0,~0.0,~0.0) & -0.31257 & ~0.83820 & & 42 & (3.5,~2.0,~0.5) & ~0.00358 & -0.00001 \\
 3 & (1.0,~0.5,~0.5) & -0.20378 & ~0.18558 & & 43 & (4.0,~0.5,~0.5) & ~0.00030 & ~0.00003 \\
 4 & (1.0,~1.0,~0.0) & ~0.27218 & -0.29230 & & 44 & (4.0,~1.0,~0.0) & ~0.00038 & -0.00118 \\
 5 & (1.5,~0.5,~0.0) & ~0.00205 & -0.05987 & & 45 & (3.0,~2.0,~2.0) & ~0.00143 & -0.00052 \\
 6 & (1.0,~1.0,~1.0) & ~0.04462 & -0.00211 & & 46 & (3.0,~2.5,~1.5) & -0.00003 & ~0.00016 \\
 7 & (1.5,~1.0,~0.5) & -0.03562 & ~0.06212 & & 47 & (3.0,~3.0,~0.0) & ~0.00358 & -0.01198 \\
 8 & (2.0,~0.0,~0.0) & -0.03068 & ~0.07661 & & 48 & (4.0,~1.0,~1.0) & -0.00122 & -0.00056 \\
 9 & (2.0,~0.5,~0.5) & ~0.01150 & ~0.00673 & & 49 & (3.5,~2.0,~1.5) & -0.00050 & -0.00105 \\
10 & (1.5,~1.5,~0.0) & ~0.08350 & -0.06621 & & 50 & (4.0,~1.5,~0.5) & -0.00057 & -0.00273 \\
11 & (2.0,~1.0,~0.0) & ~0.01215 & -0.07616 & & 51 & (3.5,~2.5,~0.0) & -0.00095 & -0.01165 \\
12 & (1.5,~1.5,~1.0) & ~0.00675 & -0.01083 & & 52 & (3.0,~3.0,~1.0) & ~0.00082 & -0.00236 \\
13 & (2.0,~1.0,~1.0) & -0.00328 & ~0.01124 & & 53 & (3.5,~2.5,~1.0) & -0.00052 & -0.00147 \\
14 & (2.5,~0.5,~0.0) & -0.00948 & ~0.01708 & & 54 & (4.0,~2.0,~0.0) & -0.00358 & ~0.00909 \\
15 & (2.0,~1.5,~0.5) & -0.02198 & ~0.04217 & & 55 & (4.5,~0.5,~0.0) & ~0.00025 & -0.00078 \\
16 & (2.5,~1.0,~0.5) & ~0.00750 & -0.00194 & & 56 & (4.0,~1.5,~1.5) & ~0.00093 & ~0.00150 \\
17 & (2.0,~2.0,~0.0) & ~0.03080 & -0.03656 & & 57 & (4.0,~2.0,~1.0) & ~0.00042 & ~0.00171 \\
18 & (2.0,~1.5,~1.5) & ~0.00063 & -0.00303 & & 58 & (3.5,~3.0,~0.5) & ~0.00032 & ~0.00660 \\
19 & (2.5,~1.5,~0.0) & ~0.00688 & -0.04823 & & 59 & (4.5,~1.0,~0.5) & -0.00093 & ~0.00041 \\
20 & (2.0,~2.0,~1.0) & ~0.00328 & -0.00889 & & 60 & (3.0,~2.5,~2.5) & -0.00005 & ~0.00002 \\
21 & (3.0,~0.0,~0.0) & ~0.00585 & -0.00193 & & 61 & (3.0,~3.0,~2.0) & -0.00003 & ~0.00006 \\
22 & (3.0,~0.5,~0.5) & -0.00287 & -0.00186 & & 62 & (3.5,~2.5,~2.0) & ~0.00232 & ~0.00016 \\
23 & (2.5,~1.5,~1.0) & -0.00118 & ~0.00076 & & 63 & (4.0,~2.5,~0.5) & ~0.00003 & -0.00114 \\
24 & (3.0,~1.0,~0.0) & -0.00775 & ~0.01594 & & 64 & (4.5,~1.5,~0.0) & ~0.00103 & -0.00023 \\
25 & (2.5,~2.0,~0.5) & -0.00937 & ~0.02320 & & 65 & (4.5,~1.5,~1.0) & -0.00047 & -0.00164 \\
26 & (3.0,~1.0,~1.0) & ~0.00363 & -0.00354 & & 66 & (3.5,~3.0,~1.5) & -0.00003 & ~0.00010 \\
27 & (3.0,~1.5,~0.5) & ~0.00617 & -0.00055 & & 67 & (4.0,~2.0,~2.0) & -0.00010 & -0.00125 \\
28 & (2.0,~2.0,~2.0) & ~0.00025 & -0.00037 & & 68 & (4.0,~2.5,~1.5) & ~0.00068 & -0.00028 \\
29 & (2.5,~2.0,~1.5) & ~0.00065 & -0.00097 & & 69 & (3.5,~3.5,~0.0) & -0.00057 & -0.00557 \\
30 & (2.5,~2.5,~0.0) & ~0.01150 & -0.02220 & & 70 & (4.5,~2.0,~0.5) & -0.00015 & -0.00206 \\
31 & (3.5,~0.5,~0.0) & ~0.00093 & ~0.00056 & & 71 & (4.0,~3.0,~0.0) & -0.00118 & -0.00410 \\
32 & (3.0,~2.0,~0.0) & ~0.00150 & -0.02508 & & 72 & (5.0,~0.0,~0.0) & ~0.00028 & -0.00067 \\
33 & (3.5,~1.0,~0.5) & ~0.00220 & -0.00129 & & 73 & (5.0,~0.5,~0.5) & ~0.00028 & -0.00019 \\
34 & (2.5,~2.5,~1.0) & -0.00178 & -0.00397 & & 74 & (3.5,~3.5,~1.0) & ~0.00003 & -0.00118 \\
35 & (3.0,~1.5,~1.5) & -0.00155 & ~0.00064 & & 75 & (4.0,~3.0,~1.0) & -0.00032 & -0.00073 \\
36 & (3.0,~2.0,~1.0) & -0.00100 & -0.00282 & & 76 & (5.0,~1.0,~0.0) & -0.00010 & -0.00014 \\
37 & (3.5,~1.5,~0.0) & -0.00553 & ~0.01155 & & 77 & (4.5,~2.5,~0.0) & ~0.00010 & ~0.00669 \\
38 & (3.5,~1.5,~1.0) & -0.00343 & ~0.00050 & & 78 & (4.5,~2.0,~1.5) & -0.00217 & ~0.00167 \\
39 & (3.0,~2.5,~0.5) & ~0.00185 & ~0.01314 & & 79 & (5.0,~1.0,~1.0) & ~0.00030 & ~0.00041 \\
40 & (4.0,~0.0,~0.0) & ~0.00000 & -0.00067 & & 80 & (3.0,~3.0,~3.0) & -0.00005 & ~0.00010 \\
\end{tabular*}
\end{ruledtabular}
\end{table*}
\endgroup

\begingroup

\begin{table*}[h]
\centering \caption{Selected three-site effective interactions. The
corresponding three-site clusters are defined by three vectors:
$R_0~\equiv~(0,0,0)$, $R_1$, $R_2$.} \label{tab:3site}
\begin{ruledtabular}
\begin{tabular*}{0.9\textwidth}{@{\extracolsep{\fill}}ccccc|cccd}
 N & \mc{1}{c}{$R_1$} &  \mc{1}{c}{$R_2$} &  \mc{1}{c}{$V^{(3)}_{(R_0,R_1,R_2)}$} & &
 N & \mc{1}{c}{$R_1$} &  \mc{1}{c}{$R_2$} &  \mc{1}{c}{$V^{(3)}_{(R_0,R_1,R_2)}$}\\
\hline
1 & (~0.5,  ~0.0,  ~0.5) & (~0.5,  ~0.5,  ~0.0) & -0.01127 & & 15 &(~0.5,  ~0.5,  ~0.0) & (~1.0,  ~1.5,  ~0.5) & ~0.00667 \\
2 & (~0.5,  ~0.5,  ~0.0) & (~1.0,  ~0.0,  ~0.0) & -0.02395 & & 16 &(~0.5,  ~0.5,  ~0.0) & (~1.5,  ~0.5,  ~1.0) & ~0.00561 \\
3 & (~0.5,  ~0.5,  ~0.0) & (~1.0,  ~0.5,  ~0.5) & ~0.04451 & & 17 &(~0.5,  ~0.5,  ~0.0) & (-0.5,  ~1.0,  ~1.5) & ~0.00213 \\
4 & (~0.5,  ~0.5,  ~0.0) & (~0.5,  ~0.5,  ~1.0) & -0.00140 & & 18 &(~1.0,  ~0.0,  ~0.0) & (~1.5,  ~1.0,  ~0.5) & ~0.00135 \\
5 & (~0.5,  ~0.5,  ~0.0) & (-0.5,  ~1.0,  ~0.5) & ~0.00470 & & 19 &(~0.5,  ~0.5,  ~0.0) & (~2.0,  ~0.0,  ~0.0) & ~0.00235 \\
6 & (~1.0,  ~0.0,  ~0.0) & (~0.5,  ~0.5,  ~1.0) & ~0.00125 & & 20 &(~0.5,  ~0.5,  ~0.0) & (~1.5,  ~1.5,  ~0.0) & -0.01867 \\
7 & (~0.5,  ~0.5,  ~1.0) & (~1.0,  -0.5,  ~0.5) & -0.00014 & & 21 &(~1.0,  ~1.0,  ~0.0) & (~2.0,  ~2.0,  ~0.0) & -0.00609 \\
8 & (~0.5,  ~0.5,  ~0.0) & (~1.0,  ~1.0,  ~0.0) & -0.04427 & & 22 &(~0.5,  ~0.5,  ~0.0) & (~2.5,  ~2.5,  ~0.0) & -0.00231 \\
9 & (~0.5,  ~0.5,  ~0.0) & (~1.0,  ~0.0,  ~1.0) & -0.00305 & & 23 &(~0.5,  ~0.5,  ~0.0) & (~2.0,  ~2.0,  ~0.0) & -0.00724 \\
10 & (~1.0,  ~0.0,  ~0.0) & (~1.0,  ~1.0,  ~0.0) & -0.00303 & & 24 &(~1.0,  ~1.0,  ~0.0) & (~2.5,  ~2.5,  ~0.0) & -0.00230 \\
11 & (~0.5,  ~0.5,  ~0.0) & (~1.5,  ~0.0,  ~0.5) & -0.00210 & & 25 &(~0.5,  ~0.5,  ~0.0) & (~3.0,  ~3.0,  ~0.0) & -0.00055 \\
12 & (~0.5,  ~0.5,  ~0.0) & (~1.5,  -0.5,  ~0.0) & -0.00181 & & 26 &(~1.0,  ~1.0,  ~0.0) & (~3.0,  ~3.0,  ~0.0) & -0.00077 \\
13 & (~0.5,  ~0.5,  ~0.0) & (~0.5,  ~0.0,  ~1.5) & -0.00324 & & 27 &(~1.5,  ~1.5,  ~0.0) & (~3.0,  ~3.0,  ~0.0) & -0.00088 \\
14 & (~0.5,  ~0.5,  ~0.0) & (~1.0,  ~1.0,  ~1.0) & -0.00687 & & 28 &(~1.5,  ~1.5,  ~0.0) & (~3.0,  ~3.0,  ~0.0) & ~0.00000 \\
\end{tabular*}
\end{ruledtabular}
\end{table*}
\endgroup

\begin{table*}[h]
\centering \caption{Selected four-site effective interactions. The
corresponding four-site clusters are defined by four vectors:
$R_0~\equiv~(0,0,0)$, $R_1$, $R_2$, $R_3$.} \label{tab:4site}
\begin{ruledtabular}
\begin{tabular*}{0.9\textwidth}{@{\extracolsep{\fill}}ccccc}
N & $R_1$ & $R_2$ & $R_3$ & $V^{(4)}_{(R_0, R_1, R_2, R_3)}$ \\
\hline
1 &(~0.5,  ~0.5,  ~0.0) & (~0.5,  ~0.0,  ~0.5) & (~0.0,  ~0.5,  ~0.5) & ~0.00418 \\
2 &(~0.5,  --0.5,  ~0.0) & (~0.5,  ~0.5,  ~0.0) & (~0.5,  ~0.0,  ~0.5) & ~0.00459 \\
3 &(~0.5,  ~0.5,  ~0.0) & (~1.0,  ~0.0,  ~0.0) & (~0.5,  --0.5,  ~0.0) & ~0.00164 \\
4 &(~0.5,  ~0.0,  ~0.5) & (~1.0,  ~0.5,  ~0.5) & (~0.5,  ~0.5,  ~0.0) & ~0.00003 \\
5 &(~1.0,  ~0.0,  ~0.0) & (~1.0,  ~0.5,  ~0.5) & (~0.5,  ~0.5,  ~0.0) & --0.00113 \\
6 &(~0.5,  ~0.5,  ~0.0) & (~0.5,  ~1.0,  ~0.5) & (~1.0,  ~0.5,  ~0.5) & ~0.00064 \\
7 &(~0.5,  ~0.5,  ~0.0) & (~1.0,  ~0.5,  ~0.5) & (~1.0,  ~0.5,  --0.5) & --0.00038 \\
8 &(~0.5,  ~0.0,  ~0.5) & (~0.5,  ~1.0,  ~0.5) & (~1.0,  ~0.5,  ~0.5) & --0.00019 \\
9 &(~0.5,  ~0.5,  ~0.0) & (~1.0,  ~1.0,  ~0.0) & (~1.0,  ~0.5,  ~0.5) & --0.00096 \\
10 &(~0.5,  ~0.5,  ~0.0) & (~1.0,  ~1.0,  ~0.0) & (~1.5,  ~1.5,  ~0.0) & ~0.00120 \\
11 &(~1.0,  ~0.0,  ~0.0) & (~1.0,  ~0.5,  ~0.5) & (~1.0,  ~0.5,  --0.5) & ~0.00022 \\
12 &(~0.0,  ~0.5,  ~0.5) & (~0.5,  ~0.0,  ~0.5) & (--0.5,  --0.5,  ~1.0) & --0.00043 \\
13 &(~1.0,  ~0.0,  ~0.0) & (~1.0,  ~0.5,  ~0.5) & (~0.5,  ~1.0,  ~0.5) & --0.00005 \\
14 &(~0.5,  --0.5,  ~0.0) & (~0.5,  --0.5,  --1.0) & (--0.5,  --1.0,  --0.5) & ~0.00005 \\
15 &(~0.5,  ~0.5,  ~0.0) & (~1.0,  ~1.0,  ~0.0) & (~1.0,  ~0.0,  ~0.0) & ~0.00032 \\
\end{tabular*}
\end{ruledtabular}
\end{table*}

\end{document}